\documentclass[reqno,oneside,12pt]{amsart}

\DeclareMathAlphabet{\mathscript}{OT1}{rsfs10}{m}{n}

\numberwithin{equation}{section}

\newif\ifShowLabels
\newdimen\theight
\def\TeXref#1{%
     \leavevmode\vadjust{\setbox0=\hbox{{\tt
          \quad\quad  {\rm \small #1}}}%
     \theight=\ht0
     \advance\theight by \dp0
     \advance\theight by \lineskip
     \kern -\theight \vbox to
          \theight{\rightline{\rlap{\box0}}%
      \vss}%
      }}%
\newcommand{\Begin}[2]%
    { \ifShowLabels \TeXref{#2} \fi \begin{#1} \label{#2} }
\newcommand{\BEGIN}[3]%
    { \ifShowLabels \TeXref{#3} \fi \begin{#1}{#2} \label{#3} }

\newcommand{\LAG}{\mathscript L}
\newcommand{\EA}{\mathcal A}
\newcommand{\EB}{\mathcal B}
\newcommand{\ED}{\mathcal D}

\begin{document}

\vspace*{-0.575in}
\begin{flushleft}
\scriptsize{GR-QC/9409039}
\end{flushleft}

\vspace*{0.5in}

\title[Sphericaly Symmetric Solutions in M\o ller's Theory]{Sphericaly 
Symmetric Solutions in M\o ller's \\ Tetrad Theory of Gravitation}
\author{F. I. Mikhail}
\address{Department of Mathematics \\ Faculty of Science \\ Ain Shams 
University \\ Cairo \\ Egypt}
\author{M. I. Wanas}
\address{Department of Astronomy \\ Faculty of Science \\ Cairo 
University \\ Cairo \\ Egypt}
\author{E. I. Lashin}
\address{Department of Physics\\ Faculty of Science \\  Ain Shams 
University \\ Cairo \\ Egypt}
\curraddr{International School for Advanced Studies (SISSA) \\ 34014 
Trieste \\ Italy}
\email{lashin@sissa.it}
\author{Ahmed Hindawi}
\address{Department of Physics\\ Faculty of Science \\  Ain Shams 
University \\ Cairo \\ Egypt}
\curraddr{Department of Physics \\ University of Pennsylvania \\ 
Philadelphia \\ PA~19104, USA}
\email{ahmed@ovrut.hep.upenn.edu}
\thanks{Published in General Relativity and Gravitation {\textbf 26} 
(1994), 869--876}

\begin{abstract}

The general solution of M\o ller's field equations in case of spherical 
symmetry is derived. The previously obtained solutions are verified as 
special cases of the general solution.

\vspace*{\baselineskip}

\noindent PACS number: 40.50.+h

\end{abstract}

\maketitle

\thispagestyle{empty}

\renewcommand{\baselinestretch}{1.2}\large\normalsize

\section{Introduction}

M\o ller \cite{Moller-78} modified General Relativity (GR) by 
constructing a new field theory using Weintzenb\"ock space. His aim was 
to get a theory free from singularities while retaining the main merits 
of GR as far as possible. For instance, the principle of general 
relativity, the principle of equivalence, and the fusion of gravity and 
mechanics are still valid. M\o ller's Theory leads to a more 
satisfactory solution to the problem of defining an energy-momentum 
complex describing the energy contents of physical systems. This problem 
has no solution in the framework of gravitational theories based on 
Riemannian space \cite{Moller-61b}. S\'aez \cite{Saez-83} generalized 
M\o ller's Theory in a very elegant and natural way into scalar tetradic 
theories of gravitation. Meyer \cite{Meyer-82} showed that M\o ller's 
Theory is a special case of Poincar\'e Gauge Theory constructed by Hehl 
et al.\ 
\cite{Hehl-et-al-80}.

In an earlier paper \cite{Mikhail-et-al-93} the authors examined this 
theory with regard to the energy-momentum complex. The authors used a 
spherically symmetric tetrad constructed by Robertson 
\cite{Robertson-32} to derive two different spherically symmetric 
solutions of M\o ller's field equations. The purpose of the present work 
is to derive the general solution of M\o ller's field equations for this 
tetrad. In Section~2 we will review briefly M\o ller's Tetrad Theory of 
Gravitation.  The structure of Weintzenb\"ock spaces with spherical 
symmetry as well as the previously derived solutions of M\o ller's field 
equations are reviewed in Section 3. The general solution of M\o ller's 
field equations is derived in Section 4. The results are discussed and 
concluded in Section 5.

\section{M\o ller's Tetrad Theory of Gravitation}

M\o ller \cite{Moller-78} constructed a gravitational theory using 
Weintzenb\"ock space for its structure. His aim was to get a theory free 
from singularities while retaining the main merits of GR as far as 
possible. In his theory the field variables are the 16 tetrad 
{components}~${e_m}^\mu$. Hereafter we use Latin indices $(mn \ldots)$ 
for vector numbers and Greek indices $(\mu\nu \ldots)$ for vector 
components. All indices run from $0$ to $3$. The metric is a derived 
quantity, given by
\Begin{equation}{1}
g^{\mu\nu} := {e_m}^\mu {e_m}^\nu.
\end{equation}
We assume imaginary values for the vector ${e_0}^\mu$ in order to have a 
Lorentz signature. We note here that, associated with any tetrad field 
${e_m}^\mu$ there is a metric field defined uniquely by \eqref{1}, while 
a given metric $g^{\mu\nu}$ does not determine the tetrad field 
completely; for any local Lorentz transformation of the tetrads 
${e_m}^\mu$ leads to a new set of tetrads which also satisfy \eqref{1}.

A central role in M\o ller's theory is played by the tensor
\begin{equation}
\gamma_{\mu\nu\sigma} := e_{m\mu}  e_{m\nu;\sigma},
\end{equation}
where the semicolon denotes covariant differentiation using the 
Christoffel symbols. This tensor has close relations to Ricci rotation 
coefficients and to torsion (cf.\ the Appendix of Ref.\ 
\cite{Meyer-82}). The tensor $\gamma_{\mu\nu\sigma}$ is invariant only 
under global Lorentz transformations. Local Lorentz invariance is lost 
in gravitational theories constructed using Weintzenb\"ock space. These 
theories admit only the weak form of the principle of equivalence. M\o 
ller considered the Lagrangian $L$ to be an invariant constructed from 
$\gamma_{\mu\nu\sigma}$ and $g_{\mu\nu}$. As he pointed out, the most 
simple possible independent expressions are
\Begin{equation}{3}
L^{(1)} := \Phi_\mu \Phi^\mu, \qquad
L^{(2)} := \gamma_{\mu\nu\sigma}\gamma^{\mu\nu\sigma}, \qquad
L^{(3)} := \gamma_{\mu\nu\sigma}\gamma^{\sigma\nu\mu},
\end{equation}
where $\Phi_\mu$ is the basic vector defined by
\begin{equation}
\Phi_\mu := {\gamma^\nu}_{\mu\nu}.
\end{equation}
These expressions $L^{(i)}$ in \eqref{3} are homogeneous quadratic 
functions in the first order derivatives of the tetrad field components.

M\o ller considered the simplest case, in which the Lagrangian $L$ is a 
linear combination of the quantities $L^{(i)}$. That is, the Lagrangian 
density is given by
\Begin{equation}{5}
\LAG_{\text{M\o ller}} := (-g)^{1/2} (\alpha_1 L^{(1)} + \alpha_2
L^{(2)} + \alpha_3 L^{(3)}),
\end{equation}
where
\begin{equation}
g := \det(g_{\mu\nu}).
\end{equation}
M\o ller chose the constants $\alpha_i$ such that his theory gives the 
same results as GR in the linear approximation of weak fields. According 
to his calculations, one can easily see that if we choose
\begin{equation}
\alpha_1 = -1, \qquad \alpha_2=\lambda, \qquad \alpha_3=1-2\lambda,
\end{equation}
with $\lambda$ equals to a free dimensionless parameter of order unity, 
the theory will be in agreement with GR to the first order of 
approximation. For $\lambda=0$, M\o ller's field equations are identical
with Einstein's equations
\Begin{equation}{8}
G_{\mu\nu} = -\kappa T_{\mu\nu}.
\end{equation}
For $\lambda\neq 0$, the field equations are given by
\Begin{align}{9}
G_{\mu\nu}+H_{\mu\nu} &= -\kappa T_{\mu\nu}, \\
F_{\mu\nu} &= 0,
\label{10}
\end{align}
where
\Begin{equation}{11}
H_{\mu\nu} := \lambda \left[ \gamma_{\alpha\beta\mu}
{\gamma^{\alpha\beta}}_{\nu} + \gamma_{\alpha\beta\mu}
{\gamma_\nu}^{\alpha\beta} + \gamma_{\alpha\beta\nu}
{\gamma_\mu}^{\alpha\beta} + g_{\mu\nu} \left( 
\gamma_{\alpha\beta\sigma}
\gamma^{\sigma\beta\alpha} - \tfrac{1}{2} \gamma_{\alpha\beta\sigma}
\gamma^{\alpha\beta\sigma} \right) \right]
\end{equation}
and
\begin{equation}
F_{\mu\nu} := \lambda \left[ \Phi_{\mu,\nu} - \Phi_{\nu,\mu} -
\Phi_{\alpha} \left( {\gamma^\alpha}_{\mu\nu} - {\gamma^\alpha}_{\nu\mu}
\right) + \smash{{\gamma_{\mu\nu}}^{\alpha}}_{;\alpha} \right].
\end{equation}
Equations \eqref{10} are independent of the free parameter $\lambda$. On 
the other hand, the term $H_{\mu\nu}$ by which equations \eqref{9} 
deviate from Einstein's field equations \eqref{8} increases with 
$\lambda$, which can be taken of order unity without destroying the 
first order agreement with Einstein's theory in case of weak fields.

\section{Spherically Symmetric Solutions in M\o ller's Theory}

The structure of Weintzenb\"ock spaces with spherical symmetry has been 
studied by Robertson \cite{Robertson-32}. In spherical polar coordinates 
the four tetrad vectors defining such structure, which admits improper 
rotations as well, can be written as
\Begin{equation}{13}
{e_m}^ \mu=
\begin{pmatrix}
A \rule{0ex}{3.2ex} & Dr & 0 & 0 \\
0 \rule{0ex}{3.2ex} & B\sin\theta\cos\phi & 
\dfrac{B}{r}\cos\theta\cos\phi
  & -\dfrac{B}{r}\dfrac{\sin\phi}{\sin\theta} \\[0.1in]
0 \rule{0ex}{3.2ex} & B\sin\theta\sin\phi & 
\dfrac{B}{r}\cos\theta\sin\phi
  & \dfrac{B}{r}\dfrac{\cos\phi}{\sin\theta} \\
0 \rule{0ex}{3.2ex} & B\cos\theta & -\dfrac{B}{r}\sin\theta & 0
\end{pmatrix},
\end{equation}
where $A,B,$ and $D$ are functions of $r$. Since one has to take the 
vector ${e_0}^\mu$ to be imaginary, in order to preserve the Lorentz 
signature for the metric, the functions $A$ and $D$ have to be taken as 
imaginary.

Using the tetrad \eqref{13} to solve M\o ller's field equations 
\eqref{9} and \eqref{10} we find that equation \eqref{10} is satisfied 
identically, and also that $H_{\mu\nu}$ as given by \eqref{11} vanishes 
identically. Thus for spherically symmetric exterior solutions, M\o 
ller's field equations are reduced to Einstein's field equations of GR, 
namely
\Begin{equation}{14}
G_{\mu\nu} = 0.
\end{equation}
Einstein tensor $G_{\mu\nu}$ may be evaluated using the Riemannian 
metric derived from \eqref{13} via the relation \eqref{11}. It is easy 
to get
\BEGIN{alignat}{2}{15}
g_{00} &= \dfrac{B^2+D^2r^2}{A^2B^2}, & \qquad
g_{10} &= g_{01} = -\dfrac{Dr}{AB^2}, \qquad
g_{11} = \dfrac{1}{B^2}, \notag \\
g_{22} &= \dfrac{r^2}{B^2}, & \qquad
g_{33} &= \dfrac{r^2\sin^2\theta}{B^2}.
\end{alignat}
The corresponding field equations \eqref{14} have given rise to 
equations (5.3)--(5.7) in Ref.\ \cite{Mikhail-et-al-93} which will not 
be repeated here.

The trivial flat space-time solution for the field equations is obtained 
by taking
\begin{equation}
A=i, \qquad B=1, \qquad D=0.
\end{equation}
A first non-trivial solution can be obtained by taking $D=0$ and solving 
for $A$ and $B$. This is the case studied by M\o ller \cite{Moller-78}, 
where he obtained the solution
\Begin{equation}{17}
A = i\frac{(1+m/2r)}{(1-m/2r)}, \qquad B = (1+m/2r)^{-2},
\qquad D = 0.
\end{equation}
A second non-trivial solution can be obtained by taking $A=i$, $B=1$, 
$D\neq 0$ and solving for $D$. In this case the resulting field 
equations can be integrated directly \cite{Mikhail-et-al-93}. The result 
is
\Begin{equation}{18}
A=i, \qquad B=1, \qquad D=i\left(\frac{2m}{r^3}\right)^{1/2},
\end{equation}
where $m$ is a constant of integration which can be identified with the 
mass of the source generating the field \cite{Mikhail-et-al-93}.

\section{General Solution of M\o ller's Field Equations}

Mikhail and Wanas \cite{Mikhail-and-Wanas-77} proposed a generalized 
field theory based on Weintzenb\"ock space, which has close formal 
similarities with M\o ller's Theory \cite{Lashin-92}. Wanas 
\cite{Wanas-85} sought spherically symmetric solutions using the tetrad 
\eqref{13}. Mazumder and Ray \cite{Mazumder-and-Ray-90} completely 
integrated the field equations of Mikhail-Wanas Theory for the tetrad 
\eqref{13} by a suitable change of variables. Due to the formal 
similarities between Mikhail-Wanas Theory and M\o ller's Theory, one 
expects a  method for solving the field equations in one theory to be 
applicable in the other.

In fact, it is clear that the tetrad \eqref{13} will have a simple $r$ 
dependence if we divide the first spatial component of every vector by 
$r$. This can be achieved by the coordinate the transformation
\begin{equation}
r \rightarrow \rho=\ln r.
\end{equation}
This transformation along with the substitutions
\begin{equation}
A = i \EA, \qquad  B = \exp(\rho) \EB, \qquad D = i \ED,
\end{equation}
where $\EA$, $\EB$, and $\ED$ are real functions of $\rho$, turn the 
tetrad \eqref{13} into the simple form
\Begin{equation}{21}
{e_m}^ \mu=
\begin{pmatrix}
i \EA \rule{0ex}{3.2ex} & i \ED & 0 & 0 \\
0 \rule{0ex}{3.2ex} & \EB \sin\theta\cos\phi & \EB \cos\theta\cos\phi
  & -\EB \dfrac{\sin\phi}{\sin\theta} \\[0.1in]
0 \rule{0ex}{3.2ex} & \EB \sin\theta\sin\phi & \EB \cos\theta\sin\phi
  & \EB \dfrac{\cos\phi}{\sin\theta} \\
0 \rule{0ex}{3.2ex} & \EB \cos\theta & -\EB \sin\theta & 0 \end{pmatrix}.
\end{equation}
The Riemannian metric associated with the tetrad \eqref{21} is given by
\BEGIN{alignat}{2}{22}
g_{00} &= \dfrac{\ED^2- \EB^2}{\EA^2 \EB^2}, & \qquad
g_{10} &= g_{01} = -\dfrac{\ED}{\EA \EB^2}, \qquad
g_{11} = \dfrac{1}{\EB^2}, \notag \\
g_{22} &= \dfrac{1}{\EB^2}, & \qquad
g_{33} &= \dfrac{\sin^2\theta}{\EB^2}.
\end{alignat}
The tetrad ${e_m}^\mu$ and the metric $g_{\mu\nu}$ as given by 
\eqref{21} and \eqref{22} do not depend on $\rho$ explicitly, in 
contrast to their explicit dependence on $r$ as given by \eqref{13} and
\eqref{15}. The corresponding field equations \eqref{14} have given rise 
to the following differential equations:
\Begin{align}{23}
3 \EB^2 \EB_\rho^2 - \EB^4   + 2 \EB \ED \EB_\rho \ED_\rho - 5 \ED^2
\EB_\rho^2 + 2 \EB \ED^2\EB_{\rho\rho} - 2 \EB^3 \EB_{\rho\rho} &= 0, 
\!\!\!\!\!\! \\
\!\!\!\!\!\!
\EA \EB^2 \EB_\rho^2 + 2 \EB^3 \EA_\rho \EB_\rho- \EA \EB^4 + 2 \EA \EB 
\ED \EB_\rho \ED_\rho - 5 \EA \ED^2 \EB_\rho^2 + 2 \EA \EB \ED^2 
\EB_{\rho\rho} &= 0, \!\!\!\!\!\! \label{24} \\
5 \EA^2 \EB \ED \EB_\rho \ED_\rho - \EA^2 \EB^2 \ED \ED_{\rho\rho} - 3 
\EA \EB \ED^2 \EA_\rho \EB_\rho - 5 \EA^2\ED^2 \EB_\rho^2 & \notag \\
+ 2 \EA^2 \EB \ED^2 \EB_{\rho\rho} - 2\EB^2\ED^2\EA_\rho^2 -\EA^2 \EB^3 
\EB_{\rho\rho} + 3 \EA \EB^2 \ED \EA_\rho \ED_\rho & \notag \\
+ \EA \EB^2 \ED^2 \EA_{\rho\rho} + 2 \EB^4 \EA_\rho^2 
- \EA^2 \EB^2 \ED_\rho^2 - \EA \EB^4 \EA_{\rho\rho} + \EA^2 \EB^2
\EB_\rho^2  & = 0, \!\!\!\!\!\!\!\!\!\!\!\!
\label{25}
\end{align}
where the subscript $\rho$ refers to differentiation with respect to 
$\rho$.

By eliminating the function $\ED$ between equations \eqref{23} and 
\eqref{24} we get
\Begin{equation}{26}
\EA \EB^2 \EB_\rho^2- \EA \EB^3 \EB_{\rho\rho} - \EB^3 \EA_\rho 
\EB_\rho=0.
\end{equation}
Equation \eqref{26} can be written in the form
\begin{equation}
\frac{\EA_\rho}{\EA} = \frac{\EB_\rho}{\EB} -
\frac{\EB_{\rho\rho}}{\EB_\rho}.
\end{equation}
This can be integrated directly. The result is
\Begin{equation}{28}
\EA = \frac{C_1 \EB}{\EB_\rho},
\end{equation}
where $C_1$ is an arbitrary constant. Equation \eqref{23} can be written 
in the form
\Begin{equation}{29}
a(\rho) \frac{d}{d\rho} \ED^2 + b(\rho) \ED^2 + c(\rho) = 0,
\end{equation}
where
\begin{equation}
\begin{split}
a(\rho) &= \EB \EB_\rho, \\
b(\rho) &= 2\EB\EB_{\rho\rho} - 5 \EB_\rho^2, \\
c(\rho) &= 3\EB^2\EB_\rho^2-\EB^4 - 2 \EB^3 \EB_{\rho\rho}.
\end{split}
\end{equation}
The general solution of equation \eqref{29} is
\Begin{equation}{33}
\ED^2=C_2 \exp\left\{ -\int \frac{b(\rho)}{a(\rho)} \, d\rho \right\}-
\exp\left\{ -\int \frac{b(\rho)}{a(\rho)} \, d\rho \right\} \times \int
\frac{c(\rho)}{a(\rho)} \exp\left\{ \int \frac{b(\rho)}{a(\rho)} \, 
d\rho \right\} \, d\rho,
\end{equation}
where $C_2$ is an arbitrary constant. The integrals in equation 
\eqref{33} can be evaluated in closed form. The result is 
\Begin{equation}{34}
\ED^2 = \frac{C_2 \EB^5}{\EB_\rho^2} - \frac{\EB^2(\EB^2 -
\EB_\rho^2)}{\EB_\rho^2}.
\end{equation}
Finally, direct substitution of $\EA$ and $\ED^2$, as given by 
\eqref{28} and \eqref{34} in equation \eqref{25}, shows that it is 
satisfied without any further restrictions on the two constants $C_1$ 
and $C_2$ or on the function $\EB$. So the general solution of 
\eqref{23}--\eqref{25} is given by \eqref{28} and \eqref{34}.

In terms of the original variables the general solution can be written 
as
\begin{equation}
\begin{split}
A &= \frac{iC_1}{1-rB'/B}, \\
D^2 &= - \frac{C_2 B^3}{r^3(1-rB'/B)^2} + \frac{BB'}{r} \frac{(2-
rB'/B)}{(1-rB'/B)^2}.
\end{split}
\end{equation}
The previously obtained solutions can be verified as special cases of 
the general solution by suitable choice of $C_1$, $C_2$, and $B$. The 
choice
\begin{alignat}{3}
C_1 &= 1, \qquad C_2 &= 2m, \qquad B &= (1+m/2r)^{-2} \\
\intertext{gives rise to the solution \eqref{17}. On the other hand,
the choice}
C_1 &= 1, \qquad C_2 &= 2m, \qquad B &= 1
\end{alignat}
gives rise to the solution \eqref{18}.

\section{Concluding Remarks}

The general solution of M\o ller's field equations for the tetrad which 
admits spherical symmetry and improper rotations has been obtained. The 
previously obtained solutions have been verified as special cases of the 
general solution.

The general solution has been found to contain an arbitrary function and 
two constants. Hence, M\o ller's field equations do not fix the tetradic 
geometry in case of spherical symmetry, up to any finite number of 
arbitrary constants. It is to be noted here that in the cosmological 
case 
\cite{Saez-and-de-Juan-84} and in the stationary axisymmetric case 
\cite{Saez-84}, it was proved that M\o ller's field equations do not fix 
the tetrad field. The general solution was not obtained in these cases. 
S\'aez \cite{Saez-83} generalized  M\o ller's Theory in a very natural 
way into scalar tetradic theories of gravitation. An important question 
in these theories is whether the field equations fix the tetradic 
geometry in the case of spherical symmetry. This question was discussed 
at length by S\'aez \cite{Saez-86}, but without giving a conclusive 
answer. In the light of the present work, it will be a major advantage 
of S\'aez's theories over M\o ller's Theory if one can answer this 
question in the affirmative. This needs more investigations before 
arriving at a final answer.

\section*{Acknowledgments}

The authors would like to express their gratitude to Dr.\ M. Melek, 
Cairo University, for his stimulating discussions. They would also like 
to thank one of their referees for his invaluable comments and 
suggestions as well as for his independent check of the validity of the 
general solution given in this paper.

\ifx\undefined\bysame
\newcommand{\bysame}{\leavevmode\hbox to3em{\hrulefill}\,}
\fi

\end{document}